\documentclass[aps,showpacs]{revtex4}
\usepackage{times}
\usepackage{graphicx}
\begin{document}
\preprint{L04-1693, accepted for publication in APL}
%%%%%%    TEXT START    %%%%%%
\title{Gallium concentration dependence of room-temperature near-bandedge luminescence in \textit{n}-type ZnO:Ga}
\author{T.~Makino}
\email[electronic mail: ]{tmakino@riken.jp}
\author{Y.~Segawa}
\affiliation{Photodynamics Research Center, RIKEN (Institute of Physical and 
Chemical Research), Aramaki aza Aoba 519-1399, Aoba-ku, Sendai 980-0845, Japan}
\author{S.~Yoshida}
\author{A.~Tsukazaki}
\author{A.~Ohtomo}
\affiliation{Institute for Materials Research, Tohoku University,
Sendai 980-8577, Japan}
\author{M.~Kawasaki}
\altaffiliation[Also at: ]{Combinatorial Materials Exploration and Technology, Tsukuba 
305-0044, Japan}
\affiliation{Institute for Materials Research, Tohoku University,
Sendai 980-8577, Japan}
\date{\today}

\begin{abstract}
We investigated the optical properties of epitaxial \textit{n}-type
ZnO films grown on lattice-matched ScAlMgO$_4$ substrates. As the
Ga doping concentration increased up to $6 \times 10^{20}$~cm$^{-3}$, the absorption edge showed a systematic blueshift, consistent
with the Burstein-Moss effect. A bright near-bandedge
photoluminescence (PL) could be observed even at room temperature,
the intensity of which increased monotonically as the doping
concentration was increased except for the highest doping level. It
indicates that nonradiative transitions dominate at a low doping
density. Both a Stokes shift and broadening in the PL band are
monotonically increasing functions of donor concentration, which was
explained in terms of potential fluctuations caused by the random
distribution of donor impurities.
\end{abstract}
% insert suggested PACS numbers in braces on next line

\maketitle

Optical properties of ZnO are currently subject of tremendous investigations, in response to the industrial demand for short-wavelength optoelectronics devices. Production of high-quality doped ZnO films is indispensable for the device application. Photoluminescence (PL) is a sensitive and non-destructive method, the results of which provide a good indicator of material quality. Impurity-doping, defect, and surface profile both have influence to its broadening, Stokes shift, and radiative efficiency. Room-temperature (RT) near-bandedge (NBE) luminescence has not been observed in donor-doped ZnO except for lightly-doped ones despite the long research history of this material as a transparent conductive window~\cite{sernelius1,ko1,lorenz1,makino19}. Indeed when ZnO:Al films were grown on lattice matched substrates, detectable NBE PL could be observed only at 5~K. As pointed out by Ko \textit{et~al.}, oxidation of the Al during the growth owing to its high reactivity may be responsible for that. On the other hand, Ga is less reactive and more resistive to oxidation. The covalent bond lengths of Ga--O is slightly smaller than that of Zn--O, which will make the deformation of the ZnO lattice small even in the case of high Ga concentration~\cite{ko1}. In this publication, we report observation of the RT NBE luminescence from ZnO:Ga epitaxial layers. The radiative efficiency, threshold energy and the linewidth of the near-band-gap optical transition are investigated as a function of doping density of Ga.

Ga-doped ZnO samples were grown by laser molecular-beam epitaxy on the (0001)-plane of a ScAlMgO$_4$ substrate. The samples were grown at temperatures of 650 to 680~$^\circ$C. The Ga doping was varied to achieve doping densities in the range of $8 \times 10^{18}$ to $6 \times 10^{20}$~cm$^{-3}$~\cite{makino8}. We used Fig.~2 of Ref.~\onlinecite{sumiya1} for the conversion from prescribed Ga concentration. The photoluminescence measurements were performed using an He-Cd laser, with emission at 325~nm. The luminescence from samples was dispersed in a 0.3~m spectrometer and detected by a charge-coupled device. Absorption was measured by using a UV/visible spectrometer (Shimadzu, UV2450)~\cite{makino19}.

Figure~1(a) shows room-temperature near-bandedge photoluminescence spectra (left-hand side) in \textit{n}-type Ga-doped ZnO samples with different doping densities. The corresponding absorption spectra (right-hand side) are also shown, and they indicate a clear blue-shift (at most $\simeq 430$~meV) related to the well-known Burstein-Moss effect~\cite{burstein1,tsmoss1}. All of the luminescence spectra displayed intense near-bandedge transitions. In order to identify the physical nature of this near-bandedge transition, comparison with absorbance data was made. In \textit{n}-doped ZnO, as shown in Fig.~1(b), an absorptive optical transition occurs from the valence band to the Fermi level or conduction band, while an emissive transition occurs from an impurity-donor band to the valence band. This is the reason for the occurrence of a Stokes shift, i.e., the luminescence peak is red-shifted from the absorption threshold. The dominant luminescence transition is therefore thought to be due to such a donor-to-free-hole recombination. Since the Stokes shifts exceed a sum of the donor and acceptor ionization energies in samples at the highest doping levels ($1.5 \times 10^{20}$ to $6 \times 10^{20}$~cm$^{-3} $), they are assigned to recombination of donor-acceptor pairs~\cite{lookp-type,tamura2}.

The intensity of the near-bandedge transition increased
markedly as the doping concentration increases. Figure~2
shows integrated intensity as a
function of doping concentration. The integrated intensity
increases by a factor of 17 as the doping density is increased
from $8 \times 10^{18}$ to $1.5 \times 10^{20}$~cm$^{-3}$ and then
finally decreases at the highest level ($6 \times 10^{20}$~cm$^{-3}$).
The relatively low intensity
at low doping concentrations is attributed to nonradiative
transitions. The lifetime of the nonradiative channel is determined
by the nonequilibrium minority carrier (hole) concentration
and the concentration of traps participating in the recombination.
In \textit{n}-type semiconductors, the trap recombination
rate is proportional to $N_Tp$, whereas the radiative
recombination rate is proportional to $np=N_Dp$. The
ratio of radiative to nonradiative recombination rates is
$N_D/N_T$~\cite{schubert2}. If $N_T$ is independent of the doping concentration,
radiative transitions increase with increase in doping concentration.
Thus, higher efficiency is expected as $N_D$ increases. This
increase in efficiency was indeed observed experimentally.
The monotonic increase in luminescence efficiency with an increase in
doping concentration also shows that luminescence killers
(deep levels) do not increase with increase in doping concentration, which
is indicative of high quality of the
epitaxial films.
In the case
of ZnO epitaxy, it has been difficult to achieve such a situation, i.e.,
impurity doping has so far induced a sizable increase in the trap center concentration~\cite{ko1,makino18,makino19}.
On the other hand, the luminescence efficiency decreases at the highest
doping concentrations. This decrease is probably attributed to
compensating native defects~\cite{calderon1}. The RT NBE PL could not
be observed in ZnO:Al films grown on the lattice-matched substrate.

We determine the energy position of the absorption edge
by taking the zero crossing of the second derivative spectrum of
the absorption coefficient.
Solid marks in Fig.~3(a) show
experimental data as a function of electron
density $n_s$ that was measured by the Hall-bar method. The experimental results are compared with the ``full'' theory of Sernelius \textit{et~al}~\cite{sernelius1}. This has been developed for polar semiconductors, taking the band gap
renormalization, Burstein-Moss effect and polaron effect into account.
A solid curve show a result which computes an energy difference between the valence band and the Fermi level. All of the quantities required to calculate this curve were taken equal as those used in Ref.~\cite{sernelius1}. The experimental data for ZnO:Ga are seen to agree well with the solid curve.

Inspection of the spectra respectively reveals that the Stokes shift energy increases from 22.4 to 396~meV and the linewidth of the transition increases from 154 to 293~meV as the doping concentration increases from $8 \times 10^{18}$ to $6 \times 10^{20}$~cm$^{-3} $. Open squares in Fig.~3(b) show the Stokes shift that is difference between the PL peak and the absorption threshold. Both the enhancement are explained in terms of potential fluctuations caused by the random distribution of doping impurities~\cite{schubert2}. It is thought that the localization of photocreated carrier due to the fluctuation determines the Stokes shift. The localization depth grows with an increase in the randomness. In a reminder of the letter, a quantitative comparison will be made only for the latter case. Randomly distributed dopants lead to unavoidable fluctuations of the doping concentration on a microscopic scale. These microscopic concentration fluctuations result in potential fluctuations. By taking the standard deviation of the potential fluctuation, the broadening of the near-bandedge transition was calculated~\cite{schubert2}, the result of which is used in this work. The full width at half-maximum (FWHM) is then given by;
%%%%%%%
\begin{eqnarray}
\Delta E_{\rm FWHM} = \frac{2e^2}{3\pi \epsilon} \sqrt{(N_D+N_A)\frac{\pi r_s}{3}}
\exp{(-3/4)} \times 2\sqrt{2\ln{2}},
\label{eq:fwhm}
\end{eqnarray}
%%%%%
where $\epsilon $ is the dielectric constant, $N_D$ ($N_A$) stands for the concentration of donors (acceptors), and $r_s$ is either the Debye or Thomas-Fermi screening radius. The factor $2\sqrt{2\ln{2}}$ accounts for the difference between the standard deviation and the FWHM of a Gaussian distribution. The other symbols used in Eq.~(\ref{eq:fwhm}) have the usual meaning.

A comparison of experimental linewidth and theoretical data is shown in Fig.~4. Since band filling is not taken into account in the
model presented here, this model is applicable only
for $N_D < n_M $, where $n_M$ is the the Mott critical density
($ \simeq 7 \times 10^{19}$~cm$^{-3}$ in ZnO)
above which the Fermi level enters into the conduction band.
The Mott density was determined from Eq.~(2) of Ref.~\onlinecite{reynolds4}.
The data for the two highest doping levels were therefore omitted.
The FWHM given by Eq.~(\ref{eq:fwhm}) and the thermal broadening given
by $1.8 kT$ are shown in Fig.~4. In addition, the total broadening
by the two uncorrelated broadening mechanisms is shown.
With an increase in concentration
from 8$ \times 10^{18}$ to $8 \times 10^{19}$~cm$^{-3} $, the FWHM became
larger, which is a similar tendency to that of the theoretical impurity broadening,
as expected.
The measured broadening is, however, quantitatively in poor agreement with the theory,
i.e., significantly larger than the
calculated broadening. Although the reason for this discrepancy is not clear,
it is probably from the contribution of the
phonon replicas superimposing on a zero-phonon luminescence band.
In the case of \textit{n}-GaN:Si, there were weak shoulders
in the Stokes sides of the main luminescence peaks~\cite{schubert2}. In ZnO:Ga, on the other hand, the intensity of a one-phonon replica could be comparable to that of a zero-phonon peak, which leads to
larger FWHMs. A further systematical study is necessary to elucidate that.

In summary, observation of the room-temperature NBE luminescence in ZnO:Ga is reported for Ga doping concentrations ranging from $8 \times 10^{18}$ to $6 \times 10^{20}$~cm$^{-3}$. A comparison of luminescence and absorption results shows that the NBE luminescence is assigned to donor-to-free-hole recombination for relatively low dopant concentrations and to donor-acceptor pairs for higher concentrations. The intensity increases monotonically with an increase in doping concentration except for the highest one, indicating the presence of luminescence killers in moderately doped ZnO. The relevancy of recombination centers is reduced at high doping concentrations. Doping yielded a band-gap widening to as large as $\sim 0.4$~eV. The sizable amounts of Stokes shift and PL broadening were explained in terms of the localization effect of photocreated carriers.
%%
%\bibliography{mybib2}

%%%\newpage 
\begin{figure} 
\includegraphics[width=0.4\textwidth]{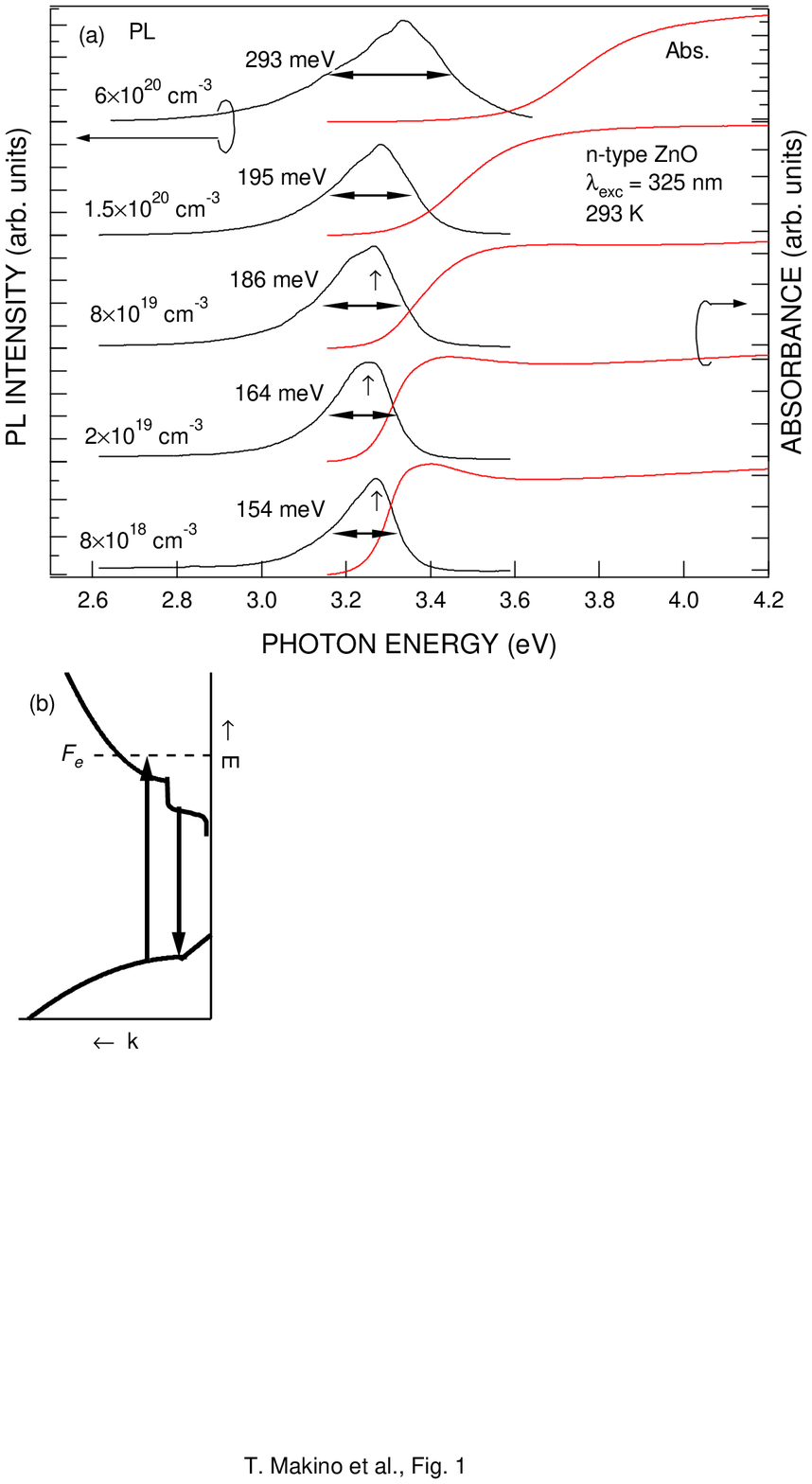}
\caption{(Color online) (a) Room-temperature photoluminescence spectra (left-hand side)
of \textit{n}-type ZnO doped at different Ga concentrations.
Also shown are the corresponding absorption spectra (right-hand side). (b) An energy diagram
of doped ZnO illustrating the corresponding optical transition
thresholds.}
\end{figure}
\begin{figure} 	
\includegraphics[width=0.4\textwidth]{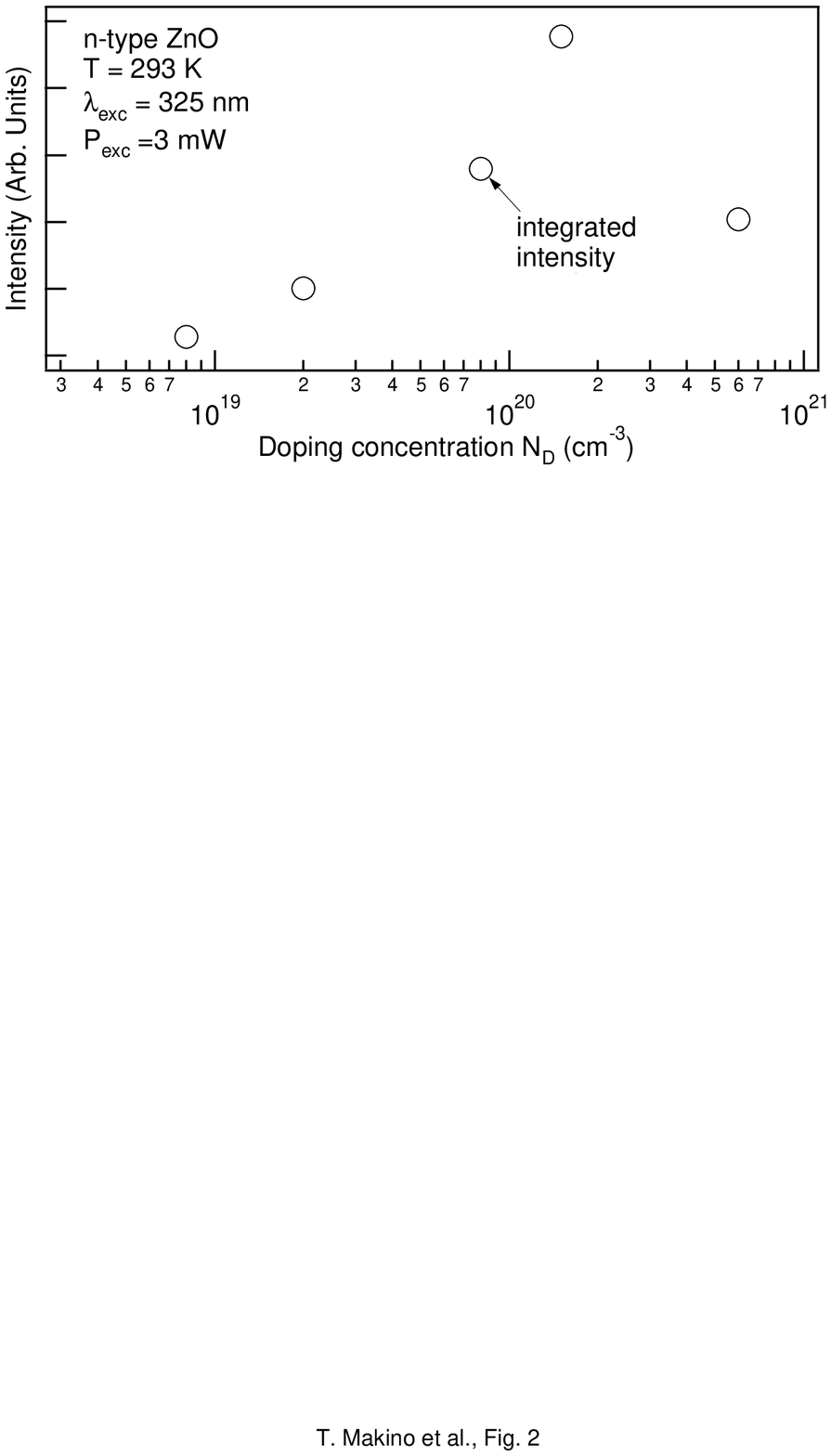}
	\caption{Integrated emission intensity of the near-bandedge transition
of \textit{n}-type ZnO as a function of the doping concentration.}
	\label{integrated}
\end{figure}
\begin{figure} 
\includegraphics[width=0.4\textwidth]{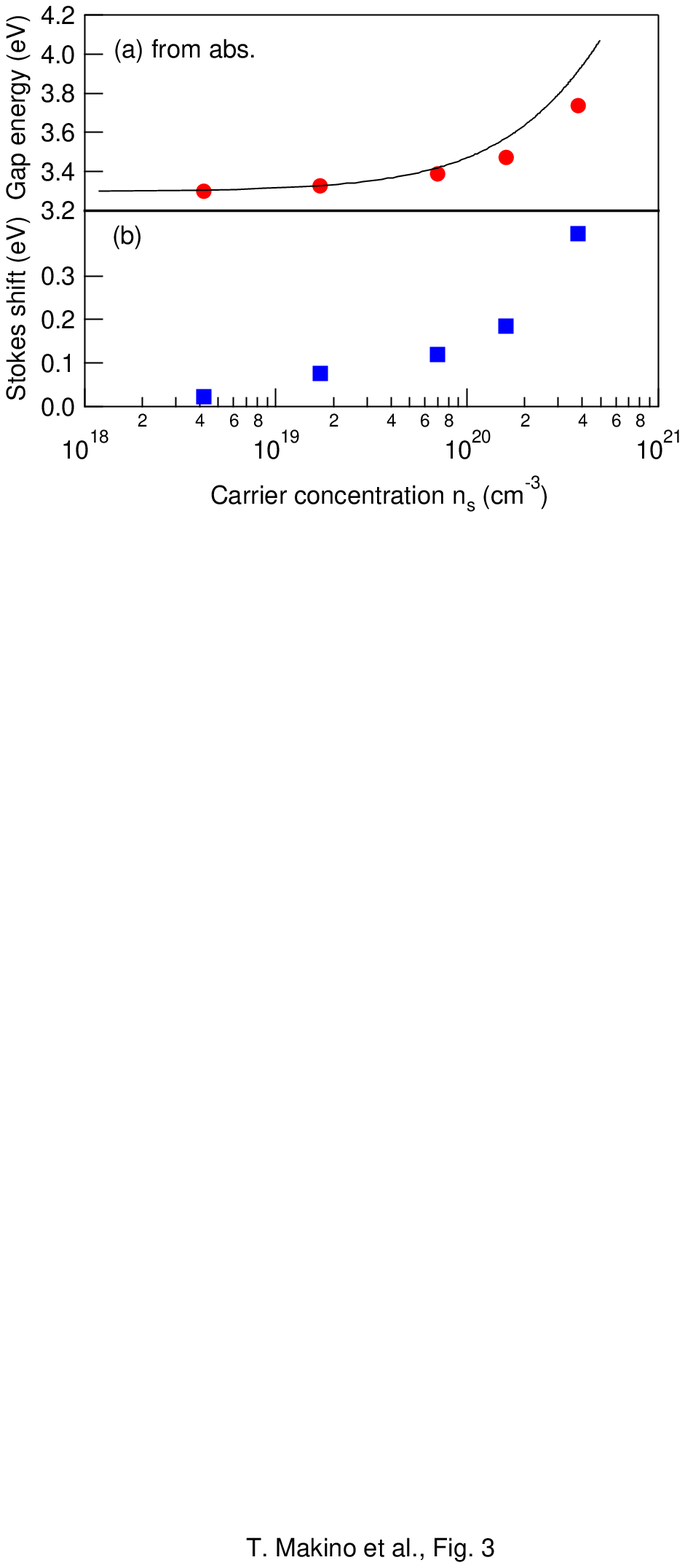}
	\caption{(Color online) (a): Optical band gap versus electron density for ZnO:Ga films as obtained from experimental data (solid circles) and a computed result (solid curves). A solid curve
refers to the theory for polar semiconductors including self-energy
shift which is cited from Ref.~\onlinecite{sernelius1}. Also shown by open squares are the PL Stokes
shifts (b).}
	\label{EnrPlot}
\end{figure}
\begin{figure} 	
\includegraphics[width=0.4\textwidth]{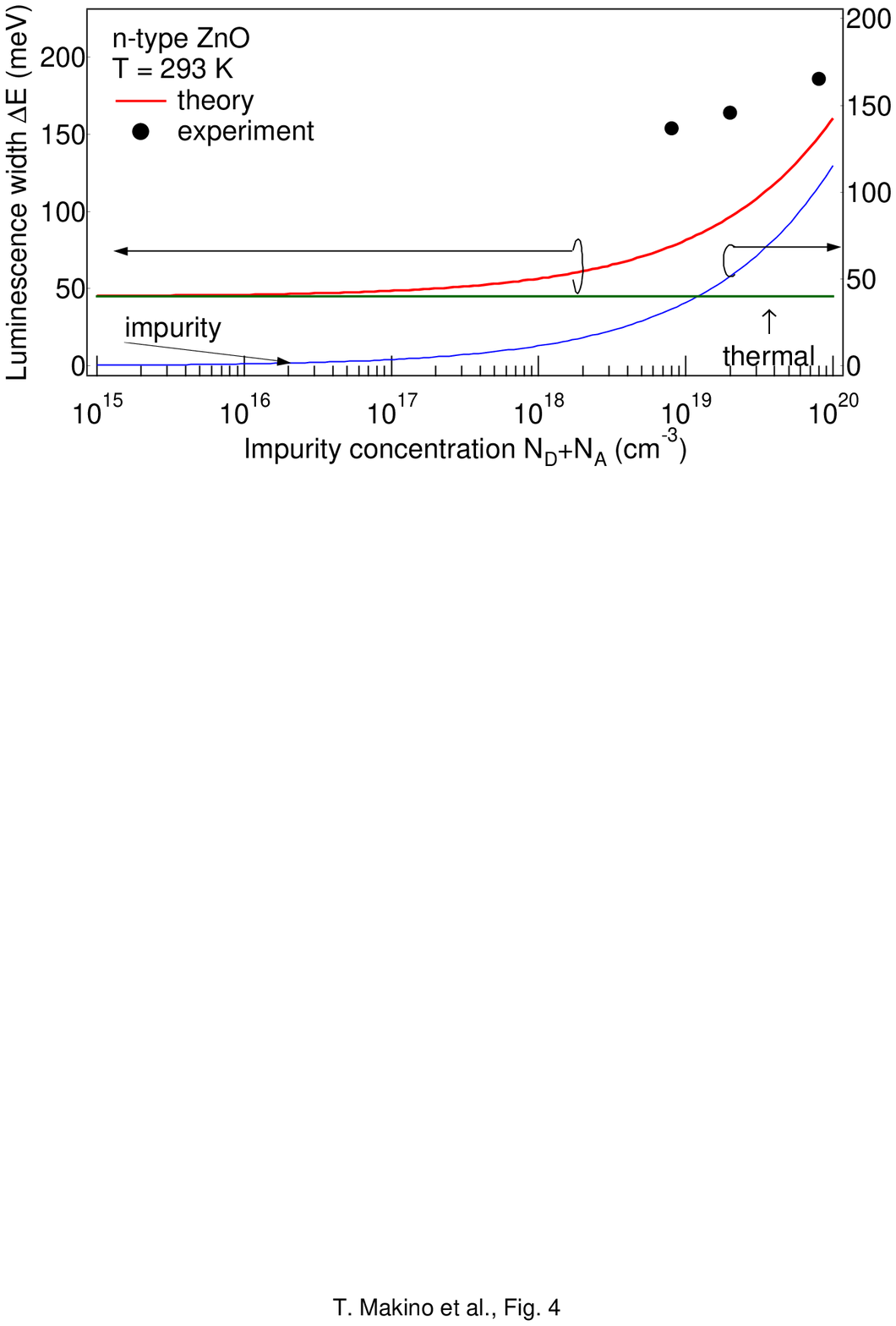}
	\caption{(Color online) Experimental linewidth of the near-bandedge transition of \textit{n}-type
ZnO as a function of the doping concentration. Also shown are the theoretical
thermal broadening and broadening due to random impurity concentration
fluctuations.}
	\label{PL}
\end{figure}

\end{document}